# Large Electron Concentration Modulation using Capacitance Enhancement in SrTiO$_3$/SmTiO$_3$ FinFETs

Amit Verma[1,2*], Kazuki Nomoto[1,2], Wan Sik Hwang[1,3], Santosh Raghavan[4], Susanne Stemmer[4] and Debdeep Jena[1,2,5]

[1]Department of Electrical Engineering, University of Notre Dame, Notre Dame, Indiana 46556, U.S.A.
[2]School of Electrical and Computer Engineering, Cornell University, Ithaca, New York 14853, U.S.A.
[3]Department of Materials Engineering, Korea Aerospace University, Goyang City, Gyeonggi-do 412791, South Korea
[4]Materials Department, University of California, Santa Barbara, California 93106, U.S.A.
[5]Department of Materials Science and Engineering, Cornell University, Ithaca, New York 14853, U.S.A.
*E-mail: averma@cornell.edu

<u>Abstract</u>: Solid-state modulation of 2-dimensional electron gases (2DEGs) with extreme (~3.3 x 10$^{14}$ cm$^{-2}$) densities corresponding to 1/2 electron per interface unit cell at complex oxide heterointerfaces (such as SrTiO$_3$/GdTiO$_3$ or SrTiO$_3$/SmTiO$_3$) is challenging because it requires enormous gate capacitances. One way to achieve large gate capacitances is by geometrical capacitance enhancement in *fin* structures. In this work, we fabricate both Au-gated planar field effect transistors (FETs) and Fin-FETs with varying fin-widths on 60nm SrTiO$_3$/5nm SmTiO$_3$ thin films grown by hybrid molecular beam epitaxy (hMBE). We find that the FinFETs exhibit higher gate capacitance compared to planar FETs. By scaling down the SrTiO$_3$/SmTiO$_3$ fin widths, we demonstrate further gate capacitance enhancement, almost twice compared to the planar FETs. In the FinFETs with narrowest fin-widths, we demonstrate a record 2DEG electron concentration modulation of ~2.4 x 10$^{14}$ cm$^{-2}$.

The recent discovery of extremely high electron density 2DEGs at the heterointerface of polar/nonpolar oxides such as LaAlO$_3$/SrTiO$_3$, GdTiO$_3$/SrTiO$_3$, SmTiO$_3$/SrTiO$_3$ has opened up the area of oxide heterostructure electronics [1-6]. These 2DEG densities are of the order of ½ electron per unit cell corresponding to ~3.3 x 10$^{14}$ cm$^{-2}$, about one order higher than 2DEG densities found in traditional semiconductor heterostructures like III-Nitrides [3,7]. Because of the high electron density, these 2DEGs are interesting for realizing high power devices and tunable infrared plasmonic devices [7]. Apart from these traditional applications, large electron concentration modulation in transition metal oxides can also enable functionally different devices based on phase-transitions [8,9]. Because of the strong electron-electron interactions in $d$-orbital derived bands of transition metal oxides, large carrier concentration changes ($\sim 10^{14}$ cm$^{-2}$ and more) can significantly change the system energy and hence the stable ground state [8,9]. A major challenge in realizing all these applications is the field-effect modulation of extreme concentrations of mobile electrons, intermediate to the concentrations in semiconductors and metals. Electron concentration modulation approaching $\sim 10^{15}$ cm$^{-2}$ has been achieved using ionic liquids in electric double layer transistors (EDLTs) [10-12]. Gating of phase transitions in complex oxides has also been demonstrated by this approach [13-16]. This EDLT approach is attractive for exploring properties of various materials as a function of carrier concentration. However, for practical applications, it is necessary to modulate carrier concentrations at high frequencies; EDLTs cannot satisfy this requirement. Because of the large inertia of ions, the high capacitance observed in EDLTs is only limited to low ($\sim 1$ kHz) modulation frequencies [11].

In traditional semiconductor devices, carrier concentration modulation at high frequencies is achieved using high-dielectric constant (high-k) gate oxides such as $HfO_2$, $Al_2O_3$ etc. One of the components in the heterointerfaces hosting extreme density 2DEGs is $SrTiO_3$, a high-k material with a large dielectric constant of $\sim 300$ at room temperature, which yields large gate capacitances [7]. Similar to other high-k gate dielectrics, $SrTiO_3$ dielectric constant remains high till THz frequencies [17]. $SrTiO_3$ is also expected to have a large breakdown field because of its large energy bandgap of 3.2 eV. Since the charge modulated in a FET is proportional to the gate capacitance and voltage applied across it, $SrTiO_3$ is expected to modulate large carrier concentrations when used as a gate dielectric. Record electron concentration modulations have indeed been demonstrated using $SrTiO_3$ gate dielectric both in heterostructures hosting 2DEGs and in MESFET devices [7,18,19].

Though $SrTiO_3$ is a promising gate dielectric for extreme electron concentration modulation, it has limitations. The dielectric constant of $SrTiO_3$ is strongly dependent on the electric field [20]. In bulk $SrTiO_3$, the dielectric "constant" decreases from 300 at zero field to $\sim 25$ at a field of $\sim 10$ MV/cm [20]. This reduction in dielectric constant limits the maximum charge modulation possible using $SrTiO_3$ as a gate dielectric. Additionally, a low dielectric constant interfacial layer (so called "dielectric dead layer") is usually present at the gate metal/ $SrTiO_3$ interface [18,19, 21-23]. This interfacial layer capacitance acts in series with $SrTiO_3$ capacitance, thus reducing the net observed gate capacitance and hence, the charge modulation in the FET. To completely modulate the extreme density 2DEGs we need approaches to reduce interfacial layers [23], increase the dielectric constant of $SrTiO_3$ [24,25], or other alternative approaches to increase the gate capacitance in the presence of these two effects. In state-of-the-art complementary metal-oxide-semiconductor (CMOS) FETs, fin shaped channels are being used for superior electrostatic gate control in nanoscale geometries [26,27]. In the fin geometry, the channel carrier concentration is modulated not only from the top but also from the sides as shown in Fig. 1(a). The fin geometry can enhance the gate capacitance even in the presence of the field-dependent dielectric constant and interfacial layer limitations. In this work, we fabricate Au-gated planar FETs and compare them with FinFETS. The FETs are fabricated on hMBE grown $SrTiO_3/SmTiO_3$ epilayers hosting extreme density 2DEGs. We demonstrate capacitance enhancement in the FinFETs compared to planar FETs. Further capacitance enhancement is observed as fin widths are scaled down. Using this capacitance enhancement, we show a record electron concentration modulation of ~2.4 x $10^{14}$ $cm^{-2}$ using $SrTiO_3$ gate dielectric.

For fabricating the FETs, 60nm $SrTiO_3$/5 nm $SmTiO_3$ thin films were grown on insulating (001) oriented $(LaAlO_3)_{0.3}(Sr_2AlTaO_6)_{0.7}$ (LSAT) substrates (Fig. 1(b)) using hMBE [28,29]. In this growth technique, Ti and O are provided using the organometallic precursor titanium isopropoxide, while Sr and Sm are provided using an effusion cell. Growth details of such heterostructures by hMBE technique have been reported earlier [3,5,28,29]. Planar FETs (plFET) and FinFETs (fFETs) with different fin widths were then fabricated on two halves of the grown sample. Fins with three different fin widths were fabricated (referred as fFET1, fFET2 and fFET3 henceforth in the order of decreasing fin-widths).

For defining the device mesas and fins, an e-beam evaporated Cr/Pt (20/30 nm) metal stack was used with Pt on top as a mesa etch mask. Mesa etching was performed in an inductively coupled plasma-reactive ion etching (ICP-RIE) system. A $BCl_3$/Ar plasma (45/5 sccm, 5 mTorr, 1000 W ICP, 97 W RIE) was used for etching to a depth of 120nm. Cr/Pt metal mask was then lifted-off in a chrome etchant. Subsequently, Al/Ni/Au (40/40/100 nm) ohmic contact metal was deposited using E-beam evaporation. A contact resistance of ~3.2 $\Omega$-mm was extracted at a drain bias of ~1V using transmission line measurements. Finally, a Schottky gate of 160nm Au was deposited by e-beam evaporation on the two samples (plFET and fFET). Prior to the Schottky metal deposition, oxygen plasma treatment (20sccm, 333 mTorr, 14 W) of the $SrTiO_3$ surface was performed in a RIE system to reduce the gate leakage and improve rectification [7]. An electron mobility of ~8.1 $cm^2$/V-sec and sheet electron concentration of

~3.09 x $10^{14}$ cm$^{-2}$ was measured for the 2DEG at 300K after the complete fabrication using Hall-effect measurements performed in a mesa isolated Van der Pauw pattern.

For measuring the FET characteristics, a Cascade probe station was used with a Keithley 4200 semiconductor characterization system. The drain current-gate voltage ($I_{ds}$-$V_{gs}$) transfer characteristics of the fabricated plFET with 50μm width and of all three fFETs (fFET1, fFET2 and fFET3) are shown in Figs. 2(a) and 2(b,c), respectively. These measurements are taken at a drain bias of 1V. Maximum reverse gate bias is limited by oxide breakdown. Since gate metal is in direct contact with the SrTiO$_3$/SmTiO$_3$ 2DEG, high-gate leakage might be expected. However, the measured gate leakage is low for all devices possibly due to the large dielectric constant of SrTiO$_3$ and large Schottky barrier height of Au [19]. As observed from Fig. 2(b), the drain current decreases in fFETs as fin widths are scaled down. By comparing the measured drain currents between plFET and fFETs at zero gate bias, we estimate the achieved fin widths in various fFET devices. Such an estimate is shown in Table I. We have been able to vary fin widths from 535 nm in fFET1 down to 192 nm in fFET3. From the Hall-effect measurements the zero gate bias 2DEG density is ~3.09 x $10^{14}$ cm$^{-2}$. Assuming that current modulation in FETs is mostly due to carrier concentration change, the calculated 2DEG charge modulation as a function of gate bias will be ~3.09 x $10^{14}$ x $|q|$/ [$I_{ds}$ ($V_{gs}$ = 0V) - $I_{ds}(V_{gs})$]/$I_{ds}$($V_{gs}$ = 0V), where $q$= -1.6x$10^{-19}$ C, the charge on an electron. This charge modulation value is shown in Fig. 3(a). The slope of charge modulation vs. gate voltage curves in Fig. 3(a) corresponds to apparent gate capacitance $C_{ap}$ obtained in the FET devices. The extracted gate capacitance values near zero bias, shown in Table I, show a clear enhancement of the gate capacitance in fFETs compared to the plFET. The gate capacitance further increases as fFET fin widths are scaled down (Fig. 3(b)).

To verify the fin and device dimensions, we performed cross-sectional and top-view scanning electron microscope (SEM) imaging of the fFETs as shown in Fig. 4. fFET1 and fFET2 fin dimensions as measured from cross-sectional SEM are ~640nm and ~350nm respectively, in close agreement to the values we estimated based on drain currents in Table I, whereas fFET3 fin seems much wider. From the SEM images, it is observed that the fFET3 fin is roughly composed of three parallel fins (Fig. 4(c), 4(f) and 4(g)). The left fin is definitely not continuous and does not contribute to the drain current. The right fin is also discontinuous in the gate-drain access region. From the cross-sectional SEM, the central fin-width is ~160nm, quite close in value to ~192nm fin-width we estimated based on drain current in fFET3. Both the current levels and SEM images therefore suggest that only the central fin is the actual channel for fFET3 device. We note that there is a ~10-20% discrepancy in fin-widths estimated from drain current and from cross-sectional SEM. This small discrepancy however has no effect on the conclusions drawn in this letter.

In all the FETs, we are modulating only part of the channel under the gate, with source/drain access regions contributing to a series parasitic resistance independent of gate bias. This access resistance roughly degrades the real capacitance $C_r$ to the apparent capacitance by the relation $C_{ap} \sim C_r R_{ch} / (R_{ch} + R_{ac})$, where $R_{ch}$ is the channel resistance under the gate and $R_{ac}$ is the total access resistance. Near zero bias, the sheet resistance in the channel and access regions is approximately the same. Therefore, the real capacitance is estimated as $C_r \sim C_{ap} (R_{ch} + R_{ac}) / R_{ch} \sim C_{ap} (L_g + L_{ac}) / L_g$, where $L_g$ is the gate length (equal to channel length under the gate) and $L_{ac} = L_{gs} + L_{gd}$ is the access region length ($L_{gs}$: gate-source access region length, $L_{gd}$: gate-drain access region length). These lengths in the plFET device are $L_g = 3.27$ μm and $L_{ac} = 3.66$ μm, while values in fFET devices are given in Table I and shown in Figs. 4(d), 4(e) and 4(f). The real capacitance values corrected for this access region effect are also given in Table I. For the plFET device which has no capacitance enhancement, the capacitance value extracted from the FET transfer characteristics is ~ 3.37 μF/cm$^2$ (Table I). To verify our extraction procedure we also performed capacitance-voltage (C-V) measurements (frequency 100 kHz, signal amplitude 30 mV) on circular Schottky diodes fabricated on the same sample along with the plFETs. The capacitance value measured at zero gate bias was ~3.35 μF/cm$^2$, quite close to the value we extracted from the transfer characteristics of the plFET. This confirms the accuracy of our capacitance extraction procedure; the access resistance corrected capacitance $C_r$ values in Table I represent the real gate capacitance in plFET and fFET devices. To calculate the capacitance enhancement in fFETs, the ratio of the fFET gate capacitance to plFET capacitance are shown in Table I. In the fFET3 device we have achieved a geometrical capacitance enhancement of ~1.94x compared to the planar FET.

In Fig. 3(b), the capacitance enhancement in FinFET devices is compared to a simple gate capacitance model based on a wrap-gate transistor geometry (Fig. 3(b) inset). The capacitance per unit length $L$ for a cylinder is $C_{cylindrical} / L = 2\pi \varepsilon_0 \varepsilon_{ox} / \ln(b / a)$, where $\varepsilon_0$ is the free space permittivity, $\varepsilon_{ox}$ is the dielectric constant of the material between the capacitor plates, and $a$ and $b$ are the inner and outer radius of the capacitor, respectively. For a parallel plate capacitor with width $W$ and length $L$, capacitance per unit length is given as $C_{parallel} / L = \varepsilon_0 \varepsilon_{ox} W / t_{ox}$, where $t_{ox}$ is the distance between the plates (equal to SrTiO$_3$ thickness in our case).  To compare the cylindrical and parallel plate capacitors, we make the assumptions that distance between the plates is same in both cases, $(b-a) \sim t_{ox}$, and width of the parallel plate capacitor is same as inner circumference of the cylindrical capacitor $2\pi a \sim W$. With these assumptions, the capacitance enhancement of cylindrical geometry will be $C_{cylindrical}/C_{parallel} \sim 2\pi/[(W/t_{ox})\ln(1+2\pi/(W/t_{ox}))]$. In Fig. 3(b), this geometrical capacitance enhancement is plotted as a function of $W/t_{ox}$ along with comparison of capacitance enhancement obtained in our FinFET devices (with $t_{ox}=60$nm and $W$ as fin-width estimated from drain currents and from cross-sectional SEM). Clearly, the capacitance enhancements we obtained from the measured transfer characteristics of fFETs and those based on the simple model agree quite well. The match is good for both the drain current and cross-sectional SEM fin-width estimations. Fig. 3(b) also suggests that further capacitance enhancements are expected with decreasing fin-width. Therefore, FinFET geometry is a very promising way to increase the gate capacitance and has the potential for complete charge modulation in SrTiO$_3$/SmTiO$_3$ extreme density 2DEGs and in other complex oxide systems. The total 2DEG density modulation achieved in plFET and fFET devices can be extracted as $\sim 3.09\text{x}10^{14}$ x$[I_{ds}(V_{gs}=+1.1\text{V}) - I_{ds}(V_{gs}=V_{br})]/I_{ds}(V_{gs}=0\text{V})$, where $3.09\text{x}10^{14}$ cm$^{-2}$ is the zero bias 2DEG density from Hall measurement and $V_{br}$ is the reverse breakdown voltage of the FET. Among all measured devices, we obtained a maximum 2DEG density modulation of $\sim 2.4 \times 10^{14}$ cm$^{-2}$ in fFET3 device with a current ON/OFF ratio of $\sim 3.2$. The measured $I_{ds}$-$V_{ds}$ characteristics of this device are shown in Fig. 5. This electron density modulation is highest ever achieved in any solid-state device using approaches other than ionic liquids.

To summarize this work, we fabricated both Au-gated planar FETs and Fin-FETs with varying fin-widths on SrTiO$_3$/SmTiO$_3$ thin films hosting extreme density 2DEGs. Compared to planar FETs, we experimentally observe a geometric gate capacitance enhancement in the fabricated FinFETs. We demonstrate further gate capacitance enhancement by scaling down the SrTiO$_3$/SmTiO$_3$ fin widths. In the fabricated FinFETs with the narrowest fin-widths, we are able to double the gate capacitance compared to the fabricated planar FETs, leading to a record 2DEG electron concentration modulation of $\sim 2.4 \times 10^{14}$ cm$^{-2}$. We believe that by further scaling down the fin-widths, complete modulation of the ½ electron per unit cell 2DEG density would be possible. We hope that improvements in large carrier concentration modulation in complex oxides using the FinFET approach will ultimately lead to reversible control of emergent phenomena in these materials.

This work was supported by the Extreme Electron Concentration Devices (EXEDE) MURI program of the Office of Naval Research (ONR) through Grant No. N00014-12-1-0976.

Table I. Capacitance enhancement in FinFETs compared to planar FET.

| Device | $I_{ds}$@$V_{gs}$=0V ($\mu A$) | Fin-width (nm) | $C_{ap}$@$V_{gs}$=0V ($\mu F/cm^2$) | $L_g$ ($\mu m$) | $L_{ac}$ ($\mu m$) | $C_r \sim C_{ap}(1+L_{ac}/L_g)$ ($\mu F/cm^2$) | $C_{FET}/C_{plFET}$ |
|---|---|---|---|---|---|---|---|
| plFET | 2260 | - | 1.59 | 3.27 | 3.66 | 3.37 | 1 |
| fFET1 | 24.2 | 535 | 2.23 | 3.42 | 2.94 | 4.15 | 1.23 |
| fFET2 | 14.7 | 325 | 2.63 | 3.32 | 2.86 | 4.89 | 1.45 |
| fFET3 | 8.7 | 192 | 4.53 | 3.31 | 1.46 | 6.53 | 1.94 |

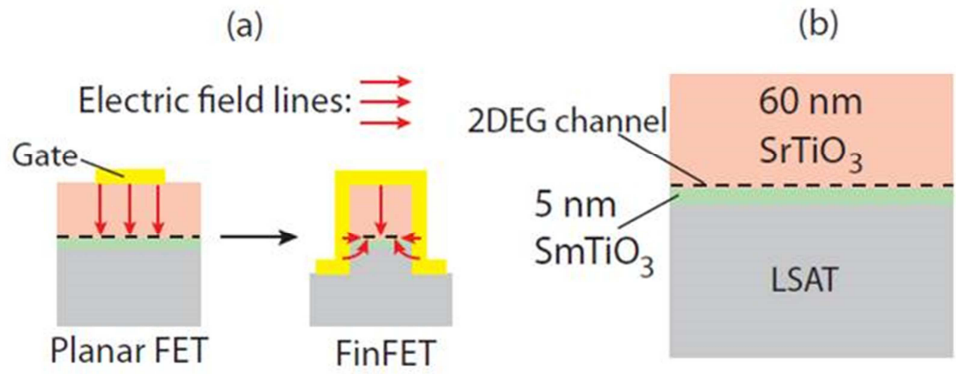

**FIG. 1**.(a) Geometrical capacitance enhancement is expected in fin geometry compared to planar FET geometry because channel is depleted from multiple sides in a FinFET, (b) Epilayer structure of the MBE grown sample used to fabricate FETs.

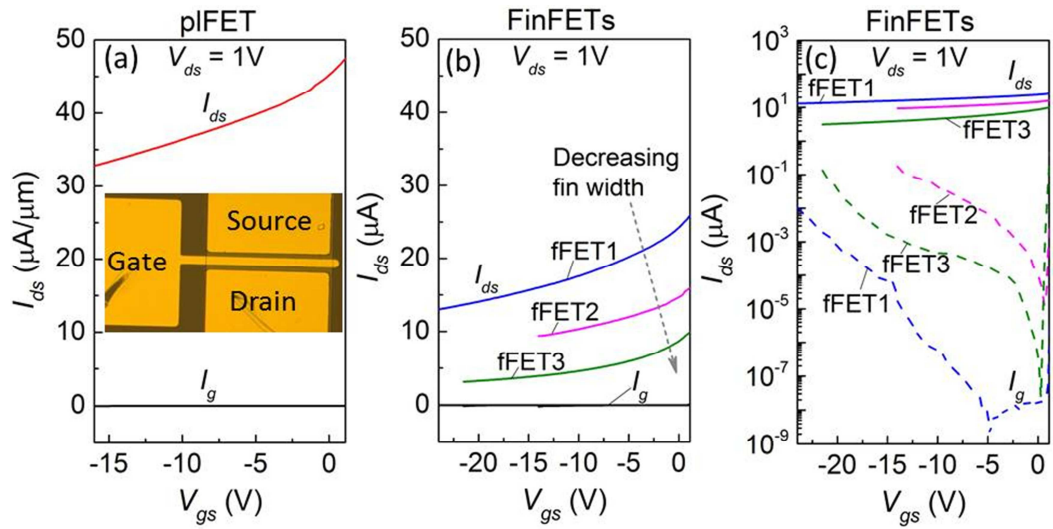

**FIG. 2.** (a) Measured $I_{ds}$-$V_{gs}$ characteristics of planar FET (Inset: Optical image of the plFET device with channel width 50 μm), (b,c) Measured $I_{ds}$-$V_{gs}$ characteristics of FinFETs showing decrease in drain current as fin-width is scaled down from fFET1 to fFET3. Gate current $I_g$ is negligible compared to drain current $I_{ds}$ for all devices.

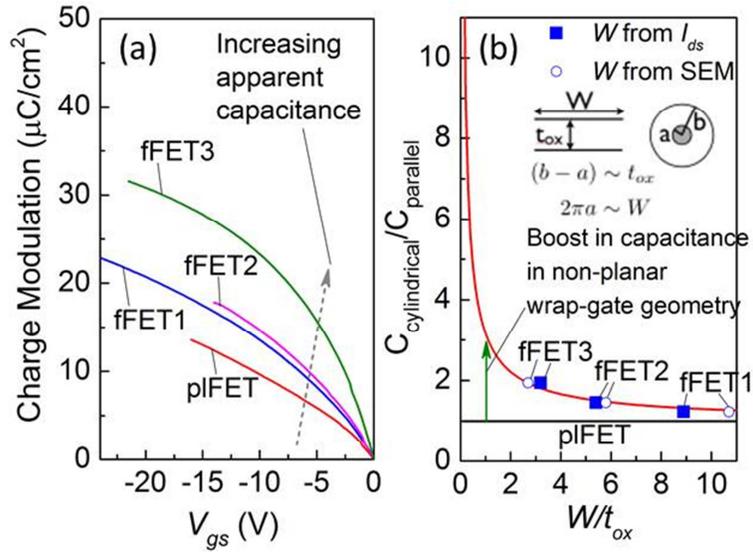

**FIG.** 3. (a) Charge modulation in all the fabricated devices as a function of gate bias; slope of the curve corresponds to apparent gate capacitance of the device; all FinFET devices show higher gate capacitance compared to planar FET; gate capacitance is observed to further increase as fin widths $W$ are scaled down from fFET1 to fFET3, (b) Capacitance enhancement in fabricated FinFET devices compared to planar FET as a function of $W/t_{ox}$, where $W$ is the fin-width (estimated from $I_{ds}$ (blue squares) and from SEM (open circles)) and $t_{ox}$ is the oxide thickness (~60nm); solid red line is the capacitance enhancement model assuming a cylindrical capacitor with difference of outer and inner radius equal to $t_{ox}$ (oxide thickness) and inner conductor circumference equal to fin-width $W$ (Inset).

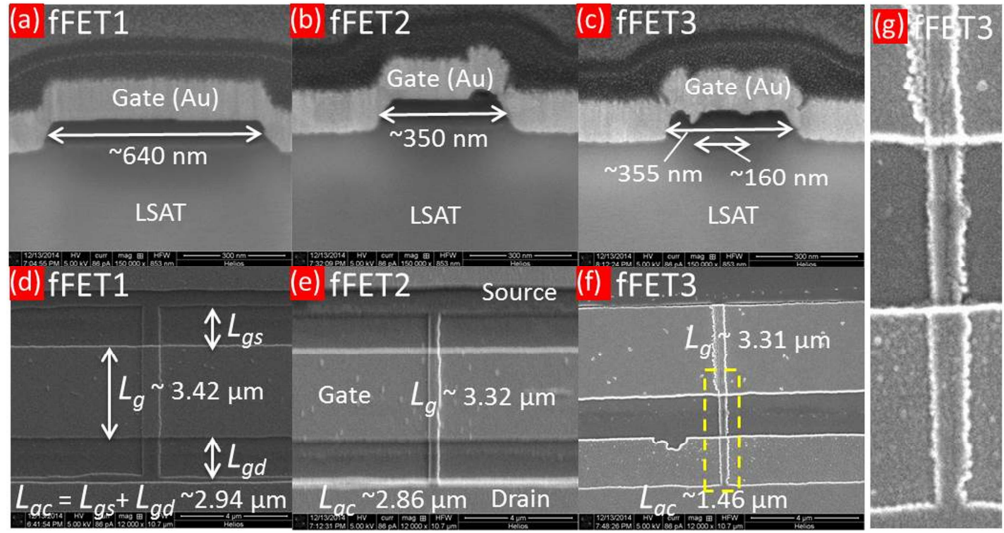

**FIG. 4.** (a,b,c) Cross-sectional and (d,e,f) top-view SEM images of fabricated FinFET devices. Scale bar is equal to 300 nm for cross-sectional images and 4 µm for top-view images. Physical fin and device dimensions are marked in the images. Highlighted gate region of fFET3 device in Fig. 4(f) is shown enlarged in Fig. 4(g), fFET3 fin is comprised of three parallel fins out of which only central fin of width~160nm is continuous and conducting.

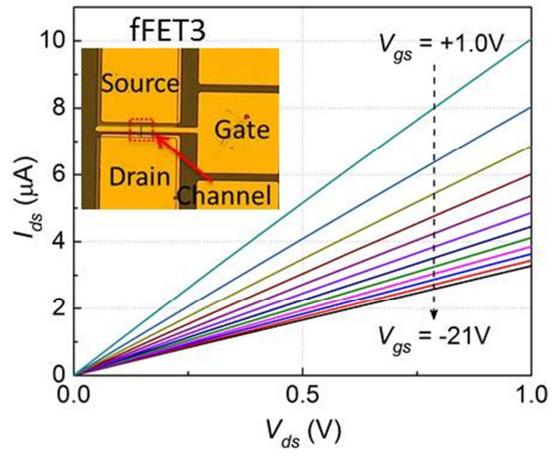

**FIG.** 5. Measured $I_{ds}$-$V_{ds}$ transistor characteristics of the fFET3 FinFET device showing 2DEG density modulation of ~2.4 x $10^{14}$ cm$^{-2}$ (Inset: Optical image of the device).